\documentclass[sigconf]{acmart}
\AtBeginDocument{%
  \providecommand\BibTeX{{%
    \normalfont B\kern-0.5em{\scshape i\kern-0.25em b}\kern-0.8em\TeX}}}

\usepackage{multirow}
\usepackage{adjustbox}
\usepackage{balance}
\usepackage{makecell}
\usepackage[utf8]{inputenc}

\usepackage{enumitem}
\usepackage{algorithm}
\usepackage{algorithmic}
\usepackage{amsmath,amsfonts}
\usepackage{array}
\usepackage{graphicx}
\usepackage{textcomp}
\usepackage[section]{placeins}

\newcolumntype{R}[2]{%
    >{\adjustbox{angle=#1,lap=\width-(#2)}\bgroup}%
    l%
    <{\egroup}%
}

\copyrightyear{2024}
\acmYear{2024}
\setcopyright{acmcopyright}\acmConference[Conference acronym 'XX]{Make sure to enter the correct conference title from your rights confirmation emai}{June 03--05, 2018}{Woodstock, NY}

\usepackage{natbib} 

\begin{document}

\title{Hierarchical Structured Neural Network:  Efficient Retrieval Scaling for Large Scale Recommendation}


\author{Kaushik Rangadurai}
\email{krangadu@meta.com}
\affiliation{%
  \institution{Meta Platforms Inc.}
  \city{Sunnyvale}
  \state{CA}
  \country{USA}
}

\author{Siyang Yuan}
\email{syyuan@meta.com}
\affiliation{%
  \institution{Meta Platforms Inc.}
  \city{Sunnyvale}
  \state{CA}
  \country{USA}
}

\author{Minhui Huang}
\email{mhhuang@meta.com}
\affiliation{%
  \institution{Meta Platforms Inc.}
  \city{Sunnyvale}
  \state{CA}
  \country{USA}
}

\author{Yiqun Liu}
\email{yiqliu@meta.com}
\affiliation{%
  \institution{Meta Platforms Inc.}
  \city{Sunnyvale}
  \state{CA}
  \country{USA}
}

\author{Golnaz Ghasemiesfeh}
\email{golnazghasemi@meta.com}
\affiliation{%
  \institution{Meta Platforms Inc.}
  \city{Sunnyvale}
  \state{CA}
  \country{USA}
}

\author{Yunchen Pu}
\email{pyc40@meta.com}
\affiliation{%
  \institution{Meta Platforms Inc.}
  \city{Sunnyvale}
  \state{CA}
  \country{USA}
}


\author{Haiyu Lu}
\email{hylu@meta.com}
\affiliation{%
  \institution{Meta Platforms Inc.}
  \city{Sunnyvale}
  \state{CA}
  \country{USA}
}

\author{Xingfeng He}
\email{xingfenghe@meta.com}
\affiliation{%
  \institution{Meta Platforms Inc.}
  \city{Sunnyvale}
  \state{CA}
  \country{USA}
}

\author{Fangzhou Xu}
\email{fxu@meta.com}
\affiliation{%
  \institution{Meta Platforms Inc.}
  \city{Sunnyvale}
  \state{CA}
  \country{USA}
}

\author{Andrew Cui}
\email{andycui97@meta.com}
\affiliation{%
  \institution{Meta Platforms Inc.}
  \city{Sunnyvale}
  \state{CA}
  \country{USA}
}

\author{Vidhoon Viswanathan}
\email{vidhoon@meta.com}
\affiliation{%
  \institution{Meta Platforms Inc.}
  \city{Sunnyvale}
  \state{CA}
  \country{USA}
}



\author{Lin Yang}
\email{ylin1@meta.com}
\affiliation{%
  \institution{Meta Platforms Inc.}
  \city{Sunnyvale}
  \state{CA}
  \country{USA}
}

\author{Liang Wang}
\email{liangwang@meta.com}
\affiliation{%
  \institution{Meta Platforms Inc.}
  \city{Sunnyvale}
  \state{CA}
  \country{USA}
}

\author{Jiyan Yang}
\email{chocjy@meta.com}
\affiliation{%
  \institution{Meta Platforms Inc.}
  \city{Sunnyvale}
  \state{CA}
  \country{USA}
}
\author{Chonglin Sun}
\email{clsun@meta.com}
\affiliation{%
  \institution{Meta Platforms Inc.}
  \city{Sunnyvale}
  \state{CA}
  \country{USA}
}

\renewcommand{\shortauthors}{Kaushik et al.}

\begin{abstract}
Retrieval, the initial stage of a recommendation system, is tasked with down-selecting items from a pool of tens of millions of candidates to a few thousands. Embedding Based Retrieval (EBR) has been a typical choice for this problem, addressing the computational demands of deep neural networks across vast item corpora. EBR utilizes Two Tower or Siamese Networks to learn representations for users and items, and employ Approximate Nearest Neighbor (ANN) search to efficiently retrieve relevant items. Despite its popularity in industry, EBR faces limitations. The Two Tower architecture, relying on a single dot product interaction, struggles to capture complex data distributions due to limited capability in learning expressive interactions between users and items. Additionally, ANN index building and representation learning for user and item are often separate, leading to inconsistencies exacerbated by representation (e.g. continuous online training) and item drift (e.g. items expired and new items added). In this paper, we introduce the Hierarchical Structured Neural Network (HSNN), an efficient deep neural network model to learn intricate user and item interactions beyond the commonly used dot product in retrieval tasks, achieving sublinear computational costs relative to corpus size. A Modular Neural Network (MoNN) is designed to maintain high expressiveness for interaction learning while ensuring efficiency. A mixture of MoNNs operate on a hierarchical item index to achieve extensive computation sharing, enabling it to scale up to large corpus size. MoNN and the hierarchical index are jointly learnt to continuously adapt to distribution shifts in both user interests and item distributions. HSNN achieves substantial improvement in offline evaluation compared to prevailing methods. HSNN has been successfully deployed in Meta’s ads recommendation system with significant online metric gains, demonstrating the effectiveness of the proposed approach in the production environment.

\end{abstract}



\maketitle

\section{Introduction}\label{sec:intro}
Machine learning plays a crucial role in recommendation systems in identifying potential user interests. To tackle the vast number of candidates per request, industry practices \cite{10.1145/3289600.3290986, 10.1145/2959100.2959190} often employ a cascade of recommendation systems with increasing computational cost. The first stage, known as the retrieval stage, narrows down millions of candidates to a few thousand.

The Retrieval stage has strict infrastructure constraints, making model architecture like Siamese Networks \cite{10.5555/2987189.2987282} a common choice. Siamese networks, or also called Two Tower, incorporate a late fusion technique where each tower outputs a fixed-sized representation. User tower uses user-only features ( e.g. user country or user click history) and the user tower is computed at query-time. Item (ad) tower uses item-only features (e.g.  content or topic of item) and item tower is computed asynchronously which is usually triggered by item updates. A single dot-product interaction between the user and item embeddings produces a logit for predicting likelihood of engagement.

Despite its popularity, the Two Tower model architecture has limitations due to its late fusion technique. This constrains the learning of sophisticated interaction between user and item, forcing features to belong to either the user (e.g. user engaged videos) or item (e.g. content of item) and prohibiting early interactions through <user, item> interactions features (e.g. user’s historical engagement of topics of this item) or model based interactions like NeuMF \cite{he2017neuralcollaborativefiltering}, DCN \cite{wang2017deepcrossnetwork} and DHEN \cite{zhang2022dhendeephierarchicalensemble} that are common in the ranking stages.

After training the Two Tower model, the user and item towers are deployed independently.  Item embeddings are computed asynchronously and used to create a vector index using systems like Manas (\cite{Manas, doshi2020lannswebscaleapproximatenearest}) and FAISS (\cite {Faiss, guo2020acceleratinglargescaleinferenceanisotropic}). Embedding Based Retrieval (EBR) \cite{Huang_2020} with approximate nearest neighbors (ANN) search \cite{johnson2017billionscale} algorithms are commonly used to retrieve top relevant candidates. Despite their popularity in the industry \cite{Manas, doshi2020lannswebscaleapproximatenearest, Faiss, guo2020acceleratinglargescaleinferenceanisotropic}, there is  inconsistency in the item embedding from index building time to scoring time. There are two main factors contributing to the dynamic nature of the item embedding distribution: \textbf{a)} The model is continuous learning and weight updates through online training, which leads to an evolving embedding distribution and \textbf{b)} The item distribution itself is also changing over time, with expired items being removed and new items being added to the index. 

NeuMF \cite{he2017neuralcollaborativefiltering}, DCN \cite{wang2017deepcrossnetwork} and DHEN \cite{zhang2022dhendeephierarchicalensemble} are advanced architectures, but their computational complexity makes them unsuitable for direct adoption in retrieval. Previous works such as Tree-based Deep Model (TDM) \cite{Zhu_2018} and Deep Retrieval (DR) \cite{gao2021deepretrievallearningretrievable} have proposed methods to enable advanced neural networks beyond two tower models in retrieval. However, these methods are limited by using the same model architecture across the tree hierarchy, which constrains further scaling up of model complexity. Moreover, the architectures proposed in TDM and DR lack expressiveness in capturing user and item interactions, failing to leverage <user, item> interaction features. Furthermore, they require iterative training to learn the hierarchy with methods such as Expectation–Maximization (EM), which can be challenging and inefficient to optimize.

We propose Hierarchical Structured Neural Network (HSNN) that can be used as a drop-in replacement to any Embedding Based Retrieval (EBR) system. HSNN brings deep neural networks to retrieval with the capability of searching through the entire corpus with both accuracy and efficiency improvements. The primary contributions of the paper include: 

\begin{itemize}
    \item We introduce HSNN which is capable of handling tens to hundreds of million items with sublinear inference cost. Concretely: \textbf{a)} Modular Neural Network (MoNN) is established to efficiently learn complex interactions among users and items with deep neural networks, \textbf{b)} A highly optimized feature transformation algorithm is introduced in MoNN to consume <user, item> interaction features. \textbf{c)} Multiple MoNNs with different complexity operate in harmony on a hierarchical index, where items falling under the same index node could share the majority of computations.
    \item We propose a gradient descent based algorithm to jointly learn the advanced neural network and hierarchical index thereby eliminating inconsistencies in neural network learning and index learning due to online training and item drift.
    \item We demonstrate the effectiveness of our proposed approach and showcase substantial performance improvements in both offline evaluation and online A/B experiments.
\end{itemize}


\section{Modular Neural Network (MoNN)}\label{sec:monn}
In this section, we introduce a new model architecture called Modular Neural Network (MoNN) and show MoNN is a more powerful model architecture leading to better performance with flexibly of operating under different infrastructure constraints.

\begin{figure*}[tb]
  \centering
  \includegraphics[width=12cm]{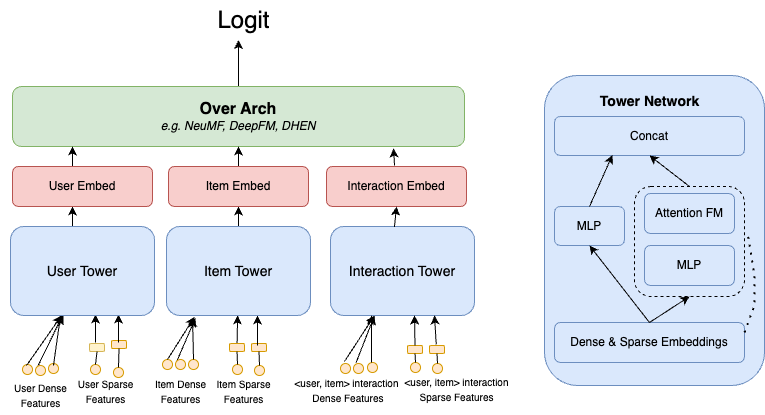}
  \caption{MoNN architecture design.}
  \label{fig:monn}
\end{figure*}

The Modular Neural Network (MoNN) enhances the learning of sophisticated user and item interactions beyond a single dot product while maintaining high efficiency. As shown in Figure~\ref{fig:monn}, it achieves this through a modularized design comprising separate modules for user representations (\textbf{User Tower}), item representations (\textbf{Item Tower}) and the interaction among user and item (\textbf{Interaction Tower}). MoNN offers high flexibility, allowing for control over the complexity of each tower to balance expressiveness and computational cost effectively.

\textbf{User Tower.} The User Tower processes user features to generate a fixed-size user embedding. These features can be dense (e.g., number of clicks by the user) or sparse (e.g., user engaged videos). Sparse features are input into an embedding table, and all feature embeddings are concatenated and fed into a Tower Network. The user tower outputs an embedding of size num\_embed$_{user}$ * dim$_{user}$ and is computed at query time. Given the user tower only needs to be computed once and shared across a vast number of items, it could scale up to very high complexity.

\textbf{Item Tower.} The Item Tower mirrors the User Tower, processing item dense (e.g. item historical click through rate) and sparse features (e.g. content of  item). Sparse features are input into an embedding table, and all feature outputs are concatenated and fed into a Tower Network, producing an embedding of size num\_embed$_{item}$ * dim$_{item}$. 

\begin{figure}[tb]
  \centering
  \includegraphics[width=8cm]{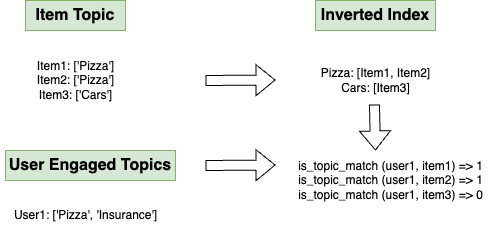}
  \caption{Inverted Index Based Interaction Features (I2IF)}
  \label{fig:i2cf}
\end{figure}

\textbf{Interaction Tower.} The Interaction Tower uses <user, item> interaction features (dense and sparse) as input. It follows a similar architecture as user tower and item tower to produce an embedding of dimension num\_embed$_{interaction}$ * dim$_{interaction}$. The Interaction Tower is computationally intensive as it runs for each pair of user and item. To minimize the computation cost for <user, item> interaction features, Inverted Index Based Interaction Features (I2IF) is introduced where an inverted index is employed for indexing item information with user information written as a query to perform efficient crossing computation. In the example shown in Figure~\ref{fig:i2cf}, the item category feature (for each item) is kept in an inverted index, and the user feature (item categories engaged by the user) is passed through the query to produce a dense interaction feature. I2IF could be applied for sparse interaction feature generation in a similar manner.

\textbf{OverArch.} Sitting atop the three underlying towers (User Tower, Item Tower and Interaction Tower), the OverArch component takes their outputs as input. It employs a DHEN style \cite{zhang2022dhendeephierarchicalensemble} model architecture to generate logits. Notably, unlike traditional Two Tower models, dim$_{item}$, dim$_{user}$ and dim$_{interaction}$ could be different and num\_embed$_{user}$, num\_embed$_{interaction}$ and num\_embed$_{item}$ could be greater than one.

\textbf{Tower Network.} The neural networks for User Tower, Item Tower and Interaction Tower follow a stacked architecture similar to EDCN \cite{10.1145/3459637.3481915}. It includes an MLP layer to capture implicit feature interactions and an AttentionFM \cite{xiao2017attentionalfactorizationmachineslearning} layer for explicit pairwise feature interactions.

\textbf{Training Setup.} The MoNN model is trained on a large training dataset with clicks and conversions as labels, impressions (non click or conversion) as negatives and additional unlabeled data used for semi-supervised learning to debias the model. A wide range of features (O(1000)) are used as input to the model. The model is optimized for multiple tasks - for e.g. click task and conversion task. The model is then trained using a multi-task cross-entropy loss:
\begin{align}
L_{sup} & = \frac{-1}{S}\sum_{i = 1}^S\sum_{t = 1}^T w_t(y_{ti}log(\hat{y}_{ti}) + (1-y_{ti}) (log(1-\hat{y}_{ti})))
\end{align}

\begin{align}
L_{unsup} = \frac{-1}{S}\sum_{i = 1}^S\sum_{t = 1}^T distil(\hat{y}_{ti}, y^{model}_{ti})
\end{align}

\begin{align}
L = L_{sup} + L_{unsup}
\end{align}

where $w_t$ is the weight for task $t$, $t$ = $1,2,...T$ , representing its importance in the final loss. $y_{ti} \in {0, 1}$ is the label for sample $i$ in task $t$. $\hat{y}_{ti}$ is the predicted value of the model for sample $i$ in task $t$ and $y^{model}_{ti}$ is the soft label generated by the MoNN model or another stronger model. $S$ is the number of samples.

\textbf{Online Training.} The retrieval model is trained through online learning where the data is ingested in a continuous stream without requiring materialization and a new model snapshot is created periodically (e.g. every few minutes) for online serving. This online training and frequent snapshot publishing, allows MoNN to consistently provide up-to-date predictions and catch-up with the item drift in the ecosystem.

\section{Hierarchical Structured Neural Network (HSNN)}\label{sec:hsnn}
While MoNN demonstrates promising ability to capture complex user and item interactions, its potential is hindered by the sheer scale of items. To overcome this limitation, we introduce HSNN, which scales up MoNN with sublinear cost in terms of corpus size. The following section delves into the details of how HSNN achieves this efficient scaling.

\textbf{Granularity of Entity Representation Matters.} A fundamental challenge in deploying state-of-the-art neural networks in the Retrieval stage is the vast cardinality of items. With thousands or tens of thousands times more items than deep neural neural networks in recommendation typically handle, this leads to:
\begin{itemize}
\item Prohibitive computational costs for advanced neural networks.
\item Exorbitant I/O access and computation costs for utilizing large amounts of features, particularly cross <user, item> features.
\end{itemize}

However, reducing the cardinality of items by $K$ times (e.g., building an item index with each index node containing K items) could theoretically enable the use of $K$ time complex models in the Retrieval without increasing serving costs.

\begin{figure}[tb]
  \centering
  \includegraphics[width=8cm]{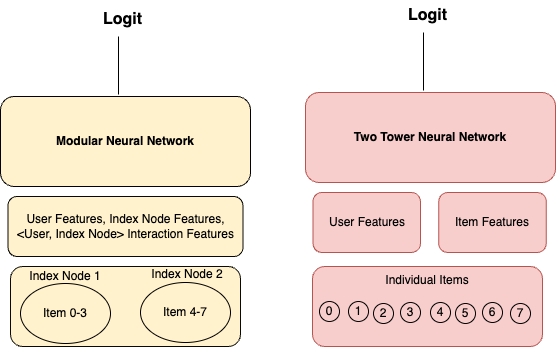}
  \caption{Neural Networks operating at different item granularities (index node vs item)}
  \label{fig:motivation}
\end{figure}

\textbf{Motivating Example.} Consider an item index with $C$ index nodes, each containing $K$ items. By assuming items within an index node share similar topics, we can reduce the complexity of candidate selection from O($K$*$C$) to O($K$+$T$), where $T$ is the number of items to return. This allows for adopting more advanced neural networks (\textbf{Model 1}) to learn sophisticated interactions between user and item (at index granularity). In contrast, Two Tower models (\textbf{Model 2}) are limited to learning basic interactions through a single dot product without any <user, item> interaction features. The comparison of Model 1 and Model 2 are shown in Figure~\ref{fig:motivation}.

While both models have pros and cons, \textbf{Model 2} captures finer-grained signals at the item level, whereas \textbf{Model 1} leverages advanced neural networks with larger signal volumes at the item index level. Ideally, a retrieval model paradigm should combine the strengths of both approaches.

\textbf{HSNN Introduction.} To harness the strengths of both models in Figure~\ref{fig:motivation}, we introduce HSNN to learn a mixture of multiple models on top of a hierarchical item index with multiple different granularities  (i.e. different granularity of item representations). By carefully designing the granularity of item index (i.e. how many items share an index node), HSNN enables the deployment of advanced ML architectures in Retrieval, capitalizing on computation sharing across entities. The coarser granularity an index is, the more computation sharing is achieved. With a hierarchical item index, HSNN can adopt a mixture of ML models with varying complexities to jointly optimize for personalization power.

HSNN consists of two primary components: (1) the hierarchical index and (2) the neural network design for each layer of the index hierarchy. In Section~\ref{sec:hsnn_model_arch}, we detail the model architecture design, while Section~\ref{sec:lti} delves into the learning of the hierarchical index.

\subsection{Model Architecture} \label{sec:hsnn_model_arch}

\begin{figure*}[tb]
  \centering
  \includegraphics[width=12cm]{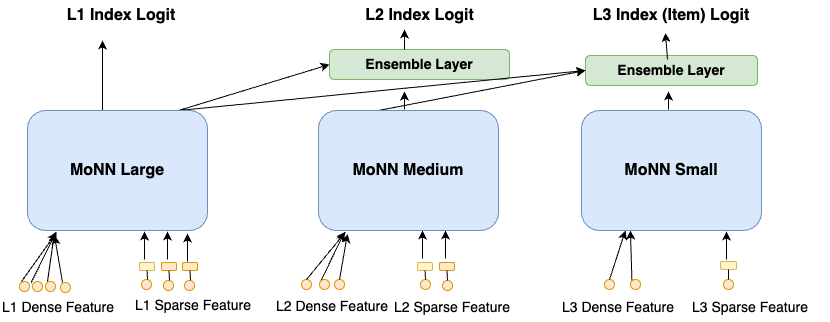}
  \caption{The model framework for 3-layer HSNN with 3 MoNNs (MoNN Large, MoNN Medium, and  and MoNN Small) operating on a 3-layer hierarchical index (L$_1$ index, L$_2$ index and L$_3$ index). L$_1$ and L$_2$ has coarse granularity of item representations, while L$_3$ layer preserving full item granularity. Each index uses sparse and dense features belonging to 3 types: user, index node and <user, index node> interaction features.}
  \label{fig:hsnn}
\end{figure*}

HSNN is a mixture of Modular Neural Networks (MoNNs) operating on top of the item hierarchy. While HSNN can be applied to N layers of hierarchy, Figure~\ref{fig:hsnn} shows a 3-layer HSNN model architecture. In this architecture, HSNN operates on a hierarchical index with three layers: L$_1$ index, L$_2$ index, and L$_3$ index. The L$_1$ index operates on the most coarse index granularity and it is able to take advantage of a MoNN model architecture with highest complexity (through computation sharing)  and a wide range of interaction features (<user, L$_1$ index node>). On the other hand, the L$_N$ index (where N=3 in this diagram) operates on the item granularity and leverages a MoNN model architecture with lowest complexity and a minimal set of <user, item> interaction features. The three MoNN modules are combined using an ensemble layer. This approach allows final prediction to leverage multiple MoNNs with different model complexity to consume  different granularity of features  resulting in a more accurate prediction.

\textbf{Features.} MoNN Small processes features at the individual item level, utilizing user features, item features and <user, item> interaction features. In contrast, MoNN Medium and MoNN Large operate at a coarser granularity (L$_2$ index and L$_1$ index respectively) and consume user features, index node features and <user, index node> interaction features where index node is replaced by a representative item for feature computation. More information on  representative item  selection is provided in Algorithm~\ref{alg:ltc_alg}.

\textbf{Loss Function.} There are N supervision losses, 1 for each layer in the hierarchy with 1 <user, item> supervision loss and (N-1) <user, index node>  supervision loss. Each of these is also calibrated to ensure that the model doesn't under-predict or over-predict. As discussed in the MoNN section, multi-task cross-entropy loss is used to optimize the model.

\begin{align}
L_{sup} & = \frac{-1}{S}\sum_{i = 1}^S\sum_{j = 1}^N\sum_{t = 1}^T w_t(y_{ti}log(\hat{y}_{ti}^{L_j}) + (1-y_{ti}) (log(1-\hat{y}_{ti}^{L_j})))
\end{align}

where $w_t$ is the weight for task $t$, $t$ = $1,2,...T$ , representing its importance in the final loss. $y_{ti} \in {0, 1}$ is the label for sample $i$ in task $t$. $\hat{y}_{ti}^{L_j}$ is the predicted value of the model for sample $i$ in task $t$ at layer $L_j$. $S$ is the number of samples and $N$ is the number of layers in HSNN.


\subsection{Learning Hierarchical Index} \label{sec:lti}

The hierarchical index can be constructed using three approaches:

\begin{enumerate}[label=(\alph*)]
\item \textbf{Exploiting existing data structures}: Leveraging the natural hierarchy of Item -> Item Creator -> Item Creator Group in the existing data structure.
\item \textbf{Clustering item embeddings}: Applying traditional clustering methods (e.g., k-means) to obtain item-index assignments based on item embeddings.
\item \textbf{Joint optimization}: Co-optimizing index assignment and the MoNN Item module.
\end{enumerate}

While approaches \textbf{a)} and \textbf{b)} have proven effective by independently learning the hierarchical index and retrieval model, we hypothesize approach \textbf{c)} can yield additional gains by co-training individual components. Therefore, we propose a Jointly Optimized indexing and MoNN \textbf{(JOIM)}, which can produce better results and generalize to learning sophisticated interactions between user and item. In the following section, we introduce a gradient descent based algorithm that learns both index assignment and representation, alongside MoNN model training. This would cover Learning to Index (LTI) for learning single layer index, learning a hierarchy of index through residual learning and the joint learning of hierarchical index and MONN.

\subsubsection{Learning To Index (LTI)}

We propose a novel Learning-to-Index (LTI) algorithm that offers three key advantages:
\begin{itemize}
    \item \textbf{Seamless integration}: Our gradient-descent-based approach can be easily incorporated into existing training infrastructure.
    \item \textbf{Lightweight and versatile}: The algorithm is designed to be computationally efficient, allowing for integration with various model architectures without significant impact on training QPS.
    \item \textbf{Gradient-friendly}: By overcoming the argmax operator in index assignment, our algorithm enables gradients to flow through, facilitating more effective optimization.
\end{itemize}

\textbf{Algorithm.} The LTI algorithm takes the item embedding from a MoNN model as input and learns coarse index node embeddings for the item by minimizing the L2 distance between them. To address the argmax operator (which identifies the closest index node to an item), it employs an attention-based method by taking the item embedding as query, learnable embeddings for index nodes as keys and values and L2 distance based attention to calculate the index embedding for them. The LTI algorithm is described in detail in Algorithm~\ref{alg:ltc_alg}. This approach calculates a soft assignment of the item to each index node based on a normalized L2 distance (line 6). It uses this to compute the item representation in the index by taking a weighted sum of index node embeddings (line 7). A key advantage of the soft assignment learning is that it allows items to belong to multiple index nodes with varying degrees of membership thereby capturing the uncertainty in the data.

\begin{algorithm}
\caption{Learning To Index (LTI)}
\begin{algorithmic}[1]
\REQUIRE MoNN model, num nodes per index (K)
\ENSURE $mapping_{item, index}$ $m$ and $representative\_tem_{index}$ $r$
\STATE Initialize embeddings$_{index}$ $\{c_{k}\}_{k=1}^K$
\FOR{each j in the batch}
\STATE Compute MoNN user embedding $u_j$ and item embedding $v_j$
\STATE $distance_{item, index}$ d(j,k) = $\|v_j - c_{k}\|^2$
\STATE $mapping_{item, index}$ m$_j = \operatorname*{argmin}_k d(j,k)$
\STATE $affinity_{item, index}$ $a_k =  \frac{e^{-\alpha * d(j,k)}}{\sum_{k'} e^{-\alpha * d(j,k')}}$
\STATE $embedding_{item , index}$ $\bar{c} = \sum_k a_k c_k$
\STATE add supervision $logloss_{index}(y, \langle u_j, \bar{c} \rangle)$
\ENDFOR
\STATE Initialize representative\_$item_{index}$
\FOR{each item j in corpora V}
\FOR{each k in K}
\STATE representative\_$item_{index, k}$ $r_k$ = j if $d(j, k)$ is closer than existing value
\ENDFOR
\ENDFOR
\STATE Publish $mapping_{item, index}$ $m$ and representative\_$item_{index}$ $r$
\end{algorithmic}
\label{alg:ltc_alg}
\end{algorithm}

\subsubsection{Residual Learning}
Inspired by Residual Quantized Variational AutoEncoder (RQ-VAE) \cite{lee2022autoregressive} \cite{zeghidour2021soundstream}, the residue between input item embedding and the corresponding index node embedding in one index layer is passed as input to the next index layer. 

Concretely, given the item embedding $v$, the $N$ level residual learning algorithm initialized the index nodes $\{c_n^k\}_{k=1, ..., K; n=1, ..., N}$ and recursively quantizes the residual vector $r_n$ at each level $n$.
\begin{align}
r_1 = v \\
r_n = r_{n-1} - \bar{c}_{n-1}
\end{align}

where $\bar{c}_{n-1}$ is the embedding$_{item, index}$ (line 7 in Equation~\ref{alg:ltc_alg}). In each level $n$, the quantized item embedding is computed as $q_n = \sum_{t=1}^{n-1}{\bar{c}_t}$. The reconstruction loss is computed as:
\begin{align}
reconstruction\ loss = \| q_N-v\|^2
\end{align}

The magnitude of residuals decreases when further moving down the hierarchy.  As a result, a more coarse index layer identifier expresses more general concepts, while a fine-grained index layer captures more detailed notions.

\subsubsection{Joint Optimization of Index and MoNN (JOIM)}

As LTI is a gradient descent approach for learning hierarchical index, it could naturally be jointly trained with MoNN. Following are some key considerations to enable it work:

\begin{itemize}
    \item \textbf{Gradient Flow}: In separate index learning, item index assignment is solely determined by item embeddings. However, joint optimization allows index assignment to be influenced by <user, item> supervision. Gradients from this supervision flow through the item tower to the LTI module, impacting item node embeddings and index assignments.
    \item \textbf{Interaction Tower}: Figure~\ref{fig:hsnn} illustrates how the MoNN module processes interaction features at both L$_1$ and L$_2$ index granularities to derive the interaction embedding. Yet it is challenging to access index level interaction features during training. To address this, we introduce a dedicated interaction tower for each index layer, utilizing only user and item features as inputs. Additionally, an auxiliary mean-squared-error loss is applied between the index and item-level interaction towers to aid in convergence.
\end{itemize}

\subsubsection{Training Optimization}

A few additional techniques are introduced to improve the training stability including softmax temperature scheduler, balanced index distribution and warmup strategy.

\textbf{Softmax Temperature Scheduler.} HSNN uses features from the representative item (line 13 from Algorithm~\ref{alg:ltc_alg}) during serving. However, during training the LTI algorithm uses a soft <item, index node> assignment. To mitigate the discrepancy, a scheduler is applied by gradually increasing the temperature (alpha), transitioning from soft assignment in the initial phase to a hard assignment later on. Small values of alpha yield a balanced distribution of the item-to-index assignment while large values of alpha result in a skewed distribution. The scheduler is based on the following function:
\begin{align}
alpha = max\_alpha * \frac{current\_iter^{exp}}{max\_iters^{exp}}    
\end{align}

\textbf{Balanced Index Distribution.} The index learning often suffers from cluster collapse, where the model utilizes only a limited subset of index nodes. A balanced index distribution is crucial to enable the use of neural network models with high complexity. We employ FLOPs regularizer to address this problem. The motivation stems from  \citeauthor{paria2020minimizing} \cite{paria2020minimizing} which penalizes the model if all items are assigned to the same index node or if the distribution of <item, index node> assignment is imbalanced. Sparsity loss is introduced to minimize the sum of squares of the mean soft assignment.  Given this is sensitive to smaller batches, data from most recent $K$ batches is pooled and the FLOPs regularizer is applied on the pooled soft assignment matrix ($K$ * batch\_size, num\_index\_nodes).

\textbf{Warmup.} A linear warm up strategy is employed for the index loss weight to gradually increase the learning rate. This approach stabilizes the model parameters and mitigates the issue of item assignment oscillating between index nodes during the initial training phase.

\section{Offline Ablation Studies}\label{sec:ablation}
\begin{table*}[tb]
\begin{tabular}{c|c|c|c|c|c}
    \toprule
    \textbf{Model Architecture} & \textbf{FLOPs} & \textbf{Eval NE ($\downarrow$)} & \textbf{Theoretical Cost} & \textbf{Infra Cost ($\downarrow$)} & \textbf{Recall ($\uparrow$)} \\
    \midrule
    Two Tower (XS) & 0.25x & baseline & M$_{XS}$ * V & 1x & 0 \\
    \hline
    MoNN Small & 1x & -0.29\% & M$_S$ * V & 2.5x & +2.4\% \\
    \hline
    MoNN Medium & 50x & -0.70\% & M$_M$ * V & 17.3x & +4.2\% \\
    \hline
    MoNN Large & 30,000x & -1.70\%  & M$_L$ * V & 24.6x & +9.4\% \\
    \hline
    MoNN XL & 200,000x & -2.20\% & M$_{XL}$ * V& 64x & +12\%  \\
    \hline
    EBR (Two Tower) & 0.25x & +0.03\% & M$_{XS}$ * I$_1$ & 0.49x & -0.1\% \\
    \hline
    \makecell{2-layer HSNN \\ (L$_1$: MoNN Small, L$_2$: TTSN)} & 1.25x & -0.23\%  & M$_S$ * I$_1$ + M$_{XS}$ * V & 1.7x & +2.2\% \\
    \hline
    \makecell{2-layer HSNN \\ (L$_1$: MoNN Medium, L$_2$: MoNN-Small)} & 51x & -0.47\% & M$_M$*I$_1$ + M$_S$ * V & 3.3x & +3.6\%  \\
    \hline
    
    \makecell{2-layer HSNN \\ (L$_1$: MoNN Large, L$_2$: MoNN Small)} & 30,001x & -0.97\% & M$_L$*I$_1$ + M$_S$ * V & 3.9x & +6\% \\
    \hline
    \makecell{3-layer HSNN \\ (L$_1$: MoNN XL, L$_2$: MoNN Large, L$_3$: MoNN Small)} & 230,0001x & -1.46\% & M$_{XL}$*I$_1$ + M$_L$*I$_2$ + M$_S$ * V& 4.5x & +10\%   \\
    \bottomrule
\end{tabular}
\vspace{2mm}
\caption{HSNN enables more complex model architectures like MoNN XL to the retrieval stage. Every scale-up brings in an order of magnitude gain. I1, I2, V are in the order of O(1,000), O(100,000), O(10,000,000) respectively. Any NE gain above 0.05\% and recall above 0.5\% is considered significant.}
\label{tab:hsnn}
\end{table*}

Ablation studies on HSNN are performed to confirm the value of key components. We use 2 offline metrics - Normalized Entropy (NE) and Recall@K to measure the performance.

\textbf{Normalized Entropy (NE).} Normalized Entropy \cite{10.1145/2648584.2648589} is equivalent to the average log loss per impression divided by what the average log loss per impression would be if a model predicted the average empirical CTR/CVR of the training data set. The lower the value, the better the model's predictions. Any number above \textbf{0.05\%} NE gain is considered to be significant.


\textbf{Recall@K.} Recall@$K$ gives a measure of how many of the relevant items are present in top $K$ out of all the relevant items, where $K$ is the number of recommendations generated for a user. A recall gain of \textbf{0.5\%} is considered to be significant.



\subsection{Advanced Model Architecture} 
This section explores whether MoNN can improve accuracy. An offline study was conducted, scoring all ads using MoNNs. We compared the Two Tower model with four MoNN variants (detailed in Section~\ref{sec:monn}) as shown in Table~\ref{tab:hsnn}. The complexity of the model is measured in Floating Point Operations (FLOPs), indicating the amount of computations needed for a single inference. All metrics reported (inference FLOPs, eval NE, infra cost and recall) are relative to the Two Tower XS model. The MoNN Small, Medium, Large, and XL models exhibit progressively larger orders of magnitude in FLOPs.

MoNN Small achieves 0.29\% NE gain over the Two Tower model architecture, highlighting the benefits of sophisticated <user, item> interaction learning provided by the Interaction Tower, Over-Arch, and <user, item> interaction features in MoNN. While large-scale MoNN variants demonstrate significant NE gains, their exponential infra cost computation makes them impractical for the retrieval stage.


\subsection{Hierarchical Structured Neural Networks (HSNN)}
Table~\ref{tab:hsnn} demonstrates the outsized gains from HSNN (up to 1.46\% NE gain and 10\% recall lift) over Two Tower architecture with magnitude less infra cost compared to vanilla MoNN, producing high return-on-investments (ROI) in Meta Ads. The relative infra cost shown in the table measures the amount of hardware capacity needed to serve the model in production with the same throughput and QPS. HSNN clearly distinguishes itself by effectively utilizing the hierarchical index with more complex MoNN architectures. In all of the HSNN experiments, the MoNN model is jointly optimized with the hierarchical index using the LTI algorithm.

For theoretical analysis, we present the cost of serving the HSNN model based on the following parameters: \textbf{I$_1$} (number of nodes in L$_1$ index), \textbf{I$_2$} (number of nodes in L$_2$ index) and \textbf{V} (number of items in the corpus), \textbf{M$_{XS}$} denotes the cost to serve Two Tower model, \textbf{M$_S$} denotes the cost to serve MoNN Small, \textbf{M$_M$} the cost to serve MoNN Medium, \textbf{M$_L$} the cost to serve MoNN Large and \textbf{M$_{XL}$} the cost to serve MoNN XL.

\subsection{Joint Optimization}
As shown in Table~\ref{tab:joim}, the Joint Optimization (\textbf{JOIM}) brings in about 0.15\% NE gain on the HSNN with separate index learning (\textbf{SIL}) with the same model architecture.  Considering the large item corpora, k-means clustering is used for its good scalability. The LTI algorithm is also compared with the \textbf{EM} variation where the MoNN model architecture and the hierarchical index are jointly learnt in an alternative way i.e.train a MoNN model with the current hierarchical index until convergence and use the converged MoNN model to update the hierarchical index structure. And LTI demonstrated better performance compared to the EM variant. Note that the infra cost for JOIM is the same as HSNN. In all the experiments, the model complexity and number of nodes is kept constant for comparison purposes.

\begin{figure}[tb]
  \centering
  \includegraphics[width=8cm]{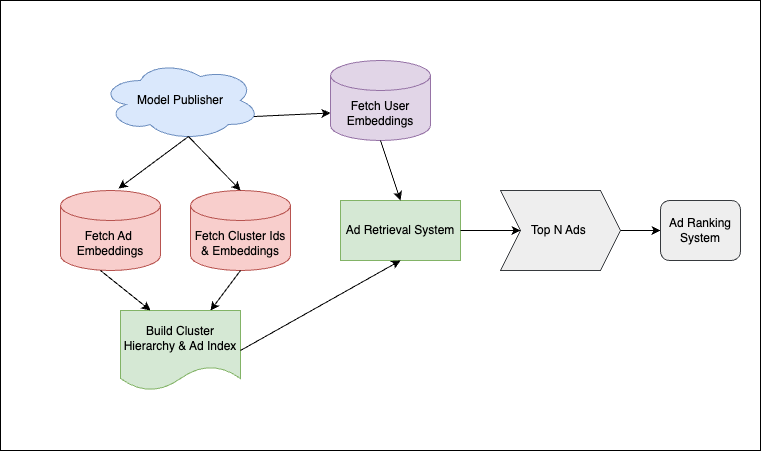}
  \caption{System architecture of the retrieval system.}
  \label{fig:hsnn_serving}
\end{figure}

\begin{table}[tb]
\begin{tabular}{c|c}
    \toprule
    \textbf{Model Architecture} & \textbf{Eval NE ($\downarrow$)} \\
    \midrule
    HSNN w/ \textbf{SIL} & baseline \\
    HSNN + JOIM w/ \textbf{EM} & -0.11\% \\
    HSNN + JOIM w/ \textbf{LTI} & -0.15\% \\
    \bottomrule
\end{tabular}
\vspace{2mm}
\caption{Ablation study on HSNN (L$_1$: MoNN Small, L$_2$: Two Tower) shows that Joint Optimization brings in 0.15\% NE gain.}
\label{tab:joim}
\vspace{-2.0em}
\end{table}

\begin{table}[tb]
\begin{tabular}{c|c}
    \toprule
    \textbf{Ablation Study} & \textbf{Eval NE ($\downarrow$)} \\
    \midrule
    HSNN (MoNN Small) & baseline \\
    - Softmax Temperature scheduler & +0.1\% \\
    - warmup strategy & +0.03\% \\
    - balanced index distribution & +0.05\% \\
    \bottomrule
\end{tabular}
\vspace{2mm}
\caption{Ablation study of various LTI components.}
\label{tab:lti}
\vspace{-2.0em}
\end{table}

As shown in Table~\ref{tab:lti}, an ablation study is performed on the various components of the LTI algorithm. The NE gains demonstrate the effectiveness of the training optimization strategies. 


\begin{table}[tb]
\begin{tabular}{c|c}
    \toprule
    \textbf{Model Architecture} & \textbf{\makecell{Online Topline \\ Metric ($\uparrow$)}} \\
    \midrule
    Two Tower & -0.21\% \\
    \hline
    MoNN Small & baseline \\
    \hline
    \makecell{HSNN + SIL \\ (L$_1$: MoNN Medium, L$_2$: MoNN Small)}  & +1.06\% \\
    \hline
    \makecell{HSNN  + JOIM \\ (L$_1$: MoNN Medium, L$_2$: MoNN Small)} & +1.22\% \\
    \hline
    \makecell{HSNN + JOIM \\ (L$_1$: MoNN Large, L$_2$: MoNN Small)} & +2.57\% \\
    \bottomrule
\end{tabular}
\vspace{2mm}
\caption{Online performance of HSNN in Meta Ads production. Any online topline metric gain above 0.1\% is considered significant.}
\label{tab:online}
\vspace{-2.0em}
\end{table}

\section{Online Experiments}\label{sec:experiments}
Figure~\ref{fig:hsnn_serving} illustrates the production deployment of the HSNN in Meta Ads. At query time, the user tower is executed to generate user embeddings, while the item tower operates asynchronously to index item embeddings. The <user, item> interaction features are generated on the fly using the  I2IF framework, feeding into the interaction tower to produce interaction embeddings in real-time. All embeddings are then passed to the OverArch and Ensemble layer to compute the logit at query time.

The HSNN framework has been widely deployed to the Meta Ads Retrieval system for over 2 years. Prior to the deployment of HSNN, we’ve launched MoNN Small architecture to production given the relatively low infra cost. As shown in Table~\ref{tab:online}, online A/B tests demonstrate that HSNN led to \textbf{2.57\%} online ads metric gains (0.1\% gain is considered as statistically significant).

\section{Related Work}\label{sec:related_work}
\textbf{Embedding Based Retrieval.} Embedding based retrieval (EBR) has been successfully applied in retrieval for search and recommendation systems \cite{que2search, rangadurai2022nxtpostuserpostrecommendations}. \cite{Huang_2020} expands this concept by integrating text, user, context, and social graph information into a unified embedding, enabling the retrieval of documents that are both relevant to the query and personalized for the user. On the systems side, many have developed or deployed approximate nearest neighbor (ANN) algorithms that can identify the top-k candidates for a given query embedding in sub-linear time \cite{Manas, Faiss, doshi2020lannswebscaleapproximatenearest}. Efficient Maximum Inner Product Search (MIPS) or ANN algorithms include tree-based methods \cite{6866199, 6809191}, locality sensitive hashing (LSH) \cite{shrivastava2014asymmetric, spring2017new}, product quantization (PQ) \cite{6678503, 5432202}, hierarchical navigable small world graphs (HNSW) \cite{malkov2018efficient}, etc.  Another research direction focuses on encoding vectors in a discrete latent space. Vector Quantized-Variational AutoEncoder (VQ-VAE) \cite{oord2018neural} proposes a simple yet powerful generative model to learn a discrete representation. HashRec \cite{kang2019candidate} and Merged-Averaged Classifiers via Hashing \cite{medini2019extreme} employ multi-index hash functions to encode items in large recommendation systems. Hierarchical quantization methods like Residual Quantized Variational AutoEncoder (RQ-VAE) \cite{zeghidour2021soundstream} and Tree-VAE \cite{manduchi2023tree} are also used to learn tree structure of a vector. However, these systems often assume a stable item vocabulary and do not account for embedding distribution shifts.

\textbf{Clustering Methods for ANN.} Clustering algorithms can be broadly categorized into hierarchical and partitional methods. Agglomerative clustering \cite{yang2016jointunsupervisedlearningdeep}, a hierarchical approach, begins with numerous small clusters and progressively merges them. Among partitioning techniques, K-means \cite{macqueen1967some} is the most well-known, aiming to minimize the sum of squared distances between data points and their nearest cluster centers.  Other approaches in this category include expectation maximization (EM) \cite{10.1111/j.2517-6161.1977.tb01600.x}, spectral clustering \cite{6321045}, and non-negative matrix factorization (NMF) \cite{10.5555/1661445.1661606} based clustering.

\textbf{Advanced Model Architectures.} Model architectures such as DCN, DHEN promote early feature interactions within the model. Due to their higher serving costs, these architectures are typically utilized in the ranking stages with a limited number of candidates to evaluate. Recently, generative retrieval has emerged as a novel paradigm for document retrieval \cite{sun2023learning, bevilacqua2022autoregressive, decao2021autoregressive, tay2022transformer, wang2023neural}. This approach parallels our work, treating items as documents and users as queries. The learned codebook in generative retrieval is akin to the item nodes in the hierarchical index proposed in this paper.

\textbf{Joint Optimization.} The line of work that most resembles our work are the joint optimized systems where the item (ad) hierarchy and the large-scale retrieval model are jointly optimized. \citeauthor{gao2021deep} \cite{gao2021deep} propose Deep Retrieval (DR), which encodes all items in a discrete latent space with an end-to-end neural network design. Deep Retrieval has constraints in terms of model architecture and features that can be utilized. \citeauthor{Zhu_2018} \cite{Zhu_2018} propose a novel tree-based method which can provide logarithmic complexity w.r.t. corpus size even with more expressive models such as deep neural networks. In this research area, tree-based methods \cite{zhu2019joint, zhuo2020learning, you2019attentionxml} are research. These approaches map each item to a leaf node within a tree structure and jointly learn an objective function for both the tree structure and model parameters.

\section{Conclusion and Next Steps}\label{sec:conclusion}
In this paper we presented a deployed retrieval model called Hierarchical Structured Neural Network (HSNN). HSNN is a powerful framework that can enable advanced neural networks to learn complex interaction among user and item features with high accuracy and efficiency. At Meta, HSNN has been successfully deployed to one of the largest Ads recommendation systems.

While hierarchical index has been heavily explored for items, it remains an important topic on how to pursue a similar strategy for users to further improve its performance. Additionally, as generative retrieval emerges, it’d be a promising direction to incorporate generative components into HSNN to further enhance its performance.

\section{Acknowledgements}\label{sec:acknowledgements}
The authors would like to thank Le Fang, Tushar Tiwari, Wei Lu, Jason Liu, Trevor Waite, Shu Yan, Alexander Petrov, Dheevatsa Mudigere, Benny Chen, GP Musumeci, Yiping Han, Bo Long, Wenlin Chen, Santanu Kolay and others who contributed, supported and collaborated with us.

\balance
\bibliographystyle{ACM-Reference-Format}
\bibliography{bibliography}

\clearpage
\appendix
\section{Appendix} \label{sec:appendix}

\subsection{MoNN Model Configurations}

 Besides FLOPs and theoretical cost  for 4 MoNN models with different complexities shown in Table~\ref{tab:hsnn}, Table~\ref{tab:hparams} provides additional information on the model configurations such as the number of features and hyper-parameters of user, item and interaction towers.

\begin{table}[tb]
\begin{tabular}{c|c|c}
    \toprule
    Model & Num Features & Tower Params  \\
    \midrule
    Two Tower Tiny & \makecell{user: O(100) \\ item: O(100) \\ interaction: 0} & \makecell{num\_embed$_{u}$=1, dim$_{u}$=40 \\ num\_embed$_{i}$=1, dim$_{i}$=40} \\
    \hline
    MoNN Small & \makecell{user: O(100) \\ item: O(100) \\ interaction: O(10)} & \makecell{num\_embed$_{u}$=8, dim$_{u}$=40 \\ num\_embed$_{i}$=1, dim$_{i}$=40 \\ num\_embed$_{ui}$=1, dim$_{ui}$=40}  \\
    \hline
    MoNN Medium & \makecell{user: O(100) \\ item: O(100) \\ interaction: O(10)} & \makecell{num\_embed$_{u}$=10, dim$_{u}$=48 \\ num\_embed$_{i}$=4, dim$_{i}$=48 \\ num\_embed$_{ui}$=1, dim$_{ui}$=48} \\
    \hline
    MoNN Large & \makecell{user: O(1000) \\ item: O(100) \\ interaction: O(100)}& \makecell{num\_embed$_{u}$=30, dim$_{u}$=48 \\ num\_embed$_{i}$=8, dim$_{i}$=48 \\ num\_embed$_{ui}$=1, dim$_{ui}$=48} \\
    \hline
    MoNN XL & \makecell{user: O(1000) \\ item: O(1000) \\ interaction: O(100)} & \makecell{num\_embed$_{u}$=80, dim$_{u}$=96 \\ num\_embed$_{i}$=4, dim$_{i}$=96 \\ num\_embed$_{ui}$=1, dim$_{ui}$=96} \\
    \bottomrule
\end{tabular}
\vspace{2mm}
\caption{MoNNs model complexities. User (u), Item (i) and Interaction (ux) are abbreviated for brevity.}
\label{tab:hparams}
\vspace{-2.0em}
\end{table}

\subsection{HSNN Serving Algorithm}

HSNN serving has two parts: the serving of the MoNN model and the inference against a hierarchical index.

A MoNN model is split into 4 parts for serving - the user tower, item tower, interaction tower and over-arch module.
\begin{itemize}
    \item The user tower takes in user features and returns the user embedding. This happens at query time.
    \item The item tower takes in item features and returns the item embedding and item-to-index mapping. An index of items is built and maintained asynchronously, with new items being immediately indexed and old items being deleted from the index.
    \item The interaction tower and over-arch model compute the interaction embedding and the logit at query time in the Ad Retrieval system.
\end{itemize}

The HSNN serving algorithm is provided in Algorithm~\ref{alg:hsnn_serving}. The HSNN serving is layer-wise. For each layer, the representative index node of the item is selected and its features are used as input to the corresponding MoNN model. The MoNN model output is passed as input to the next layer MoNN model’s linear ensemble layer to get the logit. Only the top-k nodes are selected for each layer.

\begin{algorithm}
\caption{HSNN Serving}
\begin{algorithmic}[1]
\REQUIRE MoNN Model, item-index mapping $m$
\ENSURE Top $k$ items
\FOR{each layer i in N}
\STATE Identify the index node for each item 
\STATE Evaluate MoNN Model for f(user, index-node) using the features from the representative item in the index node.
\STATE Pass the MoNN model output along with previous MoNN model outputs to the ensemble layer to get the logit.
\STATE Select top-k nodes from this layer as candidates for next layer
\ENDFOR
\STATE Publish top-k items to ranking stage.
\end{algorithmic}
\label{alg:hsnn_serving}
\end{algorithm}

\end{document}